# INTRUSION DETECTION SYSTEMS FOR IOT: OPPORTUNITIES AND CHALLENGES OFFERED BY EDGE COMPUTING AND MACHINE LEARNING

Pietro Spadaccino[1] and Francesca Cuomo[1]

[1]DIET, Department of Information Engineering, Electronics and Telecommunications, Sapienza University of Rome, 00184 Rome, Italy, (pietro.spadaccino, francesca.cuomo)@uniroma1.it

NOTE: Corresponding author: Pietro Spadaccino, pietro.spadaccino@uniroma1.it

**Abstract** – *Key components of current cybersecurity methods are the Intrusion Detection Systems (IDSs) were different techniques and architectures are applied to detect intrusions. IDSs can be based either on cross-checking monitored events with a database of known intrusion experiences, known as signature-based, or on learning the normal behavior of the system and reporting whether some anomalous events occur, named anomaly-based. This work is dedicated to survey the application of IDS to the Internet of Things (IoT) networks, where also the edge computing is used to support the IDS implementation. New challenges that arise when deploying an IDS in an edge scenario are identified and remedies are proposed. We focus on anomaly-based IDSs, showing the main techniques that can be leveraged to detect anomalies and we present machine learning techniques and their application in the context of an IDS, describing the expected advantages and disadvantages that a specific technique could cause.*

NOTE: Title, abstract and keywords must be identical to the ones submitted electronically in EDAS – Editor's Assistant. Use the command \ITUnote to achieve the appropriate formatting.

## 1. INTRODUCTION

An Intrusion Detection System (IDS) is a software or hardware component that identifies malicious actions on computer systems or networks, thus allowing security to be maintained. Host-based Intrusion Detection Systems (HIDS) target a single computer system, while Network-based Intrusion Detection Systems (NIDS) target a whole network. NIDS are devices or software components deployed in a network which analyzes the traffic generated by hosts and devices [1]. NIDSs are the focus of this work and from now on the term IDS will indicate NIDS.

The concept of IDS applied to Internet of Things (IoT) is not new and many solutions have been proposed [2][3]. Traditional IoT-oriented IDSs are placed at the device-level or at the gateway-level, as shown in Fig. 1 and in case operate by leveraging cloud computing. However, recent advances in the Edge Computing (EC) have opened new possibilities IoT that can be leveraged also from a security point of view. Indeed EC extends the Cloud Computing paradigm to the edge of the network. For example, edge computing devices, which are capable of intelligent computing, can reduce the network latency by enabling computation and storage capacity at the edge network and this is particularly significant when dealing with IoT. On the other hand, the presence of edge nodes opens new breaches which could be exploited by malicious parties for their attacks. Edge nodes could be a victim of unauthorized remote accesses or even of physical tampering, especially those nodes which are de-ployed in public areas. An attacker, gaining the control of an edge node, could alter arbitrarily all the traffic passing through it. It could selectively-forward some packets, or even injecting some new ones pretending to be a legitimate device. If IDSs were placed on device-level or gateway-level, they would not have the possibility to detect such attacks, since the intrusion takes place in a different network section. On the other hand, by deploying IDSs at the network edge (red box in Figure 1) new issues arise, which hinder the reliability of such IDSs. For these challenges to be solved, new IDSs specifically designed for the edge should be implemented.

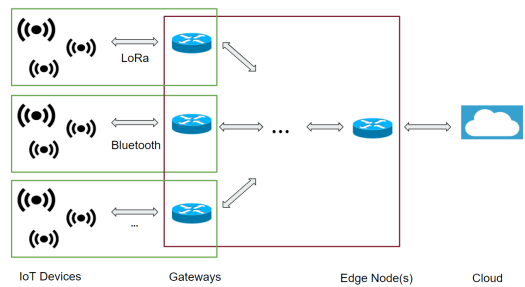

**Fig. 1** – Network architecture of an edge-enabled IoT system. Traditional IoT IDS are deployed at device-level or at gateway-level (green boxes in figure). These systems protect the network against attacks generated by some malicious IoT or non-IoT devices in the specific network. However, the network edge offers new attack surfaces to be exploited by malicious parties. IDSs could be deployed at the network edge (red box in figure). In this case new challenges arise and have to be solved, as a consequence new IDSs specifically designed for the edge should be implemented.

In this framework, the goal of this work is threefold:

- to provide a taxonomy for IDSs and to discuss their applications in the IoT field;
- to present challenges and opportunities for the implementation of IDS at the Edge;
- to survey IDSs based on Machine Learning approaches specifically tailored for IoT that can be applied at different levels of the IoT architectures.

The rest of the paper is organized as follows. In Section 2 we classify signature-based and anomaly-based IDSs. Then, in Section 3, we discuss the application of IDS to IoT environments while Section 4 specifically addresses Edge-enabled solutions. In Section 5 new challenges that arise when deploying an IDS in an edge environment are identified. We illustrate how these challenge affects existing IDS and propose possible remedies. Finally, Section 6 presents the most widely used machine learning techniques applied to IDS. For each technique, we summarize the advantages and disadvantages that the IDS should present. Conclusions are given in Section In Section 7.

## 2. INTRUSION DETECTION SYSTEM TAXONOMY

The goal of an IDS is to prevent any unauthorized access to an information system. Any access could pose a threat to information confidentiality, integrity or availability. An IDS fulfills its duty by analyzing network traffic and/or resource usage and raising an alert whether malicious activity is identified.

The IDSs can be categorized in two main families based on which strategy the system follows to detect intrusions, which can be either cross-checking monitored events with a database of known intrusion techniques or learning the normal behavior of the system and reporting whether some anomalous event occurs. These strategies are named *signature-based* and *anomaly-based*, respectively.

### 2.1 Signature-based IDS

Signature-based IDSs (SIDSs) are a class of systems that leverage a database of "signatures" of known attacks. Signatures of the current activities are extracted and matching methods and/or protocol conformance checks are then used to compare these signatures to the ones in the database. If a matching is found an alarm is triggered. They can operate both in online mode, directly monitoring the hosts and raising alarms in real-time, and offline mode, where logs of the system activities are analyzed. This class of IDS is also known in the literature as Misuse Detection or Knowledge-based Detection [4]. The extraction of traffic signatures may be a cumbersome and lengthy task to carry on, depending on which and how many traffic "features" are considered.

Indeed signatures are often manually-crafted by experts having detailed knowledge of the exploits that the system is supposed to capture. Christian Kreibich et al. [5] proposed a system for automatic generation of malicious traffic signatures. They extended the open source honeypot `honeyd` [6] with a subsystem inspecting traffic at various levels in the protocol hierarchy, and integrating it with existing IDS. The first approaches of SIDS analyze single network packets and match them with the rules in the database. However, it may be necessary to extract and match signatures spanning over multiple packets, thus building matching rules considering previous observed packets. Meiners et al. [7] proposed a signature matching method based on finite state machines. They have also implemented a hardware-based regular expression matching approach using small Ternary Content Addressable Memories chips. Lin et al. [8] proposed an algorithm performing matching based on algorithms such as Backward Hashing and Aho-Corasick. They partitioned the signatures to be examined so that they could benefit from choosing a matching algorithm instead of another based on the signatures' features. They have also implemented the system in the HIDS CalmAV. Sheikh [9] et al. proposed a signature-based Intrusion Detection System for IoT environments. The system is composed by four sub-systems in cascade. First signatures of known attack types are extracted and stored in a databse, which should be updated frequently to increase the accuracy of the detection. Then sessions are merged extracting features and a novel pattern matching algorithm is applied to compare the incoming data with the known signatures. Finally, the system produces logs to be inspected by the system administrator, who can check the current and historical status of it.

However, signature-based IDS cannot cope with zero-day attacks, which are attacks whose signature is not in the IDS's database. The rising rate of zero-day attacks [10] makes less effective the overall performance of a signature-based IDS. For this reason anomaly-based IDSs were developed, a new class of IDSs which model the nominal behaviour of a computer system and then reports any significant deviation from the baseline.

### 2.2 Anomaly-based IDS

Anomaly-based IDS (AIDS) were developed to overcome the limitations of a signature-based IDS. AIDS usually have a training phase, during which they build a model of the nominal behavior of the system. When the IDS is deployed, it monitors computer hosts and compares their behavior with the nominal one. When a significant deviation between the hosts' behavior and the model is observed, the IDS may raise an alert. Potentially, this strategy gives an anomaly-based IDS the capability to capture zero-days attacks, since it does not perform any matching between the current hosts behavior and attack signatures in a database. Another advantage of an anomaly-based IDS, is that it is difficult for an at-tacker to understand the normal behavior of a target host without doing transactions with it, since communicating with a target would likely make the IDS to raise an alert [11], [12]. Moreover, anomaly-based IDS could be exploited not just for security purposes, but also as a system analysis tool. If the IDS reports anomalies, it means that something is working differently from the baseline conditions, which can be an indication of not only an intrusion, but can also show the presence of a bug in the devices logic. A major limitation of an AIDS, is the higher rate of false positives when compared to a SIDS. Indeed, during operation, the targeted system can slightly or drastically change behavior without any intrusion taking place, and if an AIDS is not aware of this possibility it can raise false alerts.

Anomaly-based IDSs can be divided into three subcategories based on their modelling and detection techniques: statics-based, knowledge-based and machine learning-based. Statistics-based and machine learning-based IDS builds a model of the normal hosts' behavior, while the knowledge-based focus on capturing anomalies on system data such as network protocol or pattern in data exchange provided by the system administrators. The use of different IDSs belonging to different classes is not exclusive, since an IDS category can capture certain attacks which could be undetected by other IDSs using other detection techniques, providing multi-tier security.

### 2.2.1 Statistics-based AIDS

During the learning phase, an IDS based on statistical techniques builds a probability distribution model of the computing system during its nominal behavior. The model is built by taking measurements of different parameters and events taking place in the computing system. When the IDS is deployed, it evaluates the probability of all the monitored events of the system, and raises alerts on low probability events. The simplest strategy to build the statistical model is the so-called "Univariate" strategy, and consists in considering each measurement independently from the others. An evolution of it is the "Multivariate" strategy, which consists in identifying correlations and relationships between two or more measurements. Ye et al. [13] proposed a hybrid univariate and multivariate system, by building profiles of each measurement individually, and then discover multivariate correlations to decrease the false positives rate. Tan et al. [14] developed a system able to detect DoS attacks following the anomaly strategy. The system learns through multivariate correlation analysis (MCA) the normal traffic patterns of the system, and raises alert in case of DoS with high accuracy. When dealing with a high number of measurements, using multivariate statistics techniques on the raw data may produce a high level of noise. To overcome this problem, systems in [15], [16], [17], [18] used principal component analysis (PCA), a statistical technique which is used to reduce the dimensions of input vectors, before applying standard multivariate statistical techniques.

Another family of statistics-based techniques to detect anomalies takes advantage of time series data techniques [19], which have also been applied to network anomaly detection [20]. When calculating the probability of an event occurring also the time is considered, and an alert is raised if an event is unlikely to have happened in a specific time. Viinikka et al. [21] exploits time series techniques by aggregating individual alerts into an alert flow, and examining it as a whole. This has the benefit to perform a more precise multivariate analysis and to lower the false positive rate of alerts, since irrelevant alerts can be discovered at flow level and not be raised. Qingtao et al. [22] proposed a system focused on detecting abrupt-change anomalies of the computing system. They used the Auto-Regressive (AR) process to model the data, and then performed a sequential hypothesis testing to determine the presence of an anomaly. Zhao et al. [23] exploited techniques to mine frequent patterns in network traffic, and applied time-decay factors to differentiate between newer and older patterns. This strategy helps AIDS to update its system baseline, making the IDS to cope with highly dynamical behavior of users. When developing an AIDS, and especially an AIDS exploiting time series data, attention must be given to data seasonality. Seasonality is the presence of variations in the data, which occur periodically in a course of months, days or even hours. It can be caused by "human" factors like holidays and work-hours or can be also influenced by other factors like weather, depending on the application. Reddy et al. [24] proposed an algorithm to detect outliers in seasonality-affect time series data using a double pass of Gaussian Mixture Models (GMMs). During the learning phase, they divide the time into seasonal time bins, GMMs are trained and outliers are removed from data. To improve performance, another set of GMMs are built on the cleaned data. Finally this second set of GMMs is used to carry out the final anomaly detection.

### 2.2.2 Knowledge-based AIDS

Knowledge-based AIDS falls in the category of the so-called expert systems. These systems leverage a knowledge source which represents the legitimate traffic signature. Every event that differs from this profile is treated as an anomaly. This knowledge is, most of the times, hand-crafted, and could contain rules about the nominal traffic patterns of the systems as well as Finite State Machines (FSM) applied to internet protocols such as IP, TCP, HTTP, etc. to ensure the compliance of the host to the aforementioned protocols. Walkinshaw et al. [25] have applied FSMs to the whole network traffic, representing the activities of the system by states and transitions. The produced FSM represents the nominal behavior of the system, and any deviation from is considered an attack. Studnia et al. [26] developed a system based on description language defining the characteristics of an attack. Knowledge-based AIDSs leverage a precise model of the whole computing system. This model enables them to reach a low rate of false positives when compared to other solutions. On the other hand, it may be a difficult and cumbersome task, if not unfeasible, to hand-craft a model of the system. This model should be flexible enough to overcome dynamic changes in the system behavior, and most of the time this family of AIDS is applied only on predictable traffic sources [26]. Ensuring protocol compliance via a FSM could be a hard task, since we have to model our state machine on top of the targeted protocol, which can be complex. Moreover, there could be the risk that the implementation is bugged. Also, if the implementation used by the IDS is the same to the one used by the hosts (e.g. open protocol stacks implemented in Linux kernel), there is the risk to duplicate those bugs. In this case, the IDS is unable to catch protocol violations and may introduce vulnerabilities in the whole Intrusion Detection System.

### 2.2.3 Machine Learning-based AIDS

Machine Learning (ML) has been extensively applied in the field of cybersecurity [27]. Many of the specialized branches of ML have been exploited to develop an AIDS, including Data Mining [28], Deep Learning [29] [30], Deep Reinforcement Learning [31] and lately Adversarial Learning [32]. ML-based AIDSs leverage Machine Learning models to automatically learn a representation of the normal conditions of the computing system.

When designing a ML system, the first step is to identify the features of the data to be analyzed, and an IDS makes no exception [33]. Preliminary works are focused on evaluating the goodness of traffic features, by using publicly available datasets and applying baseline ML algorithms. Works from Khraisat [34], Bajaj [35] and Elhag [36] evaluate the importance of dataset features via Information Gain (IG), Correlation Attribute Evaluation and by applying genetic-fuzzy rule mining methods. By exploiting this evaluation, they clean out features that bring low IG or carry the same information of another feature. They then apply algorithms such as C4.5 Decision Tree, Naïve Bayes, NB-Tree, Multi-Layer Perceptron, SVM, and k-means Clustering. Other techniques used for IDS feature selections include Principal Component Analysis (PCA) [37], [38] and Genetic Algorithms (GA) [39].

A Machine Learning model can be trained with or without ground-truth labels. The learning techniques take the name Supervised Learning and Unsupervised Learning. The Supervised Learning strategy consists in giving as input to the ML algorithm the data alongside with their true labels (anomaly / not anomaly). Techniques that leverage this training strategy include Support Vector Machines (SVMs), Artificial Neural Networks (ANNs), Decision Trees, etc. Since the ML model knows anomalous events, having carried on the training procedure on both normal and abnormal events, an AIDS trained in a supervised way usually presents a high accuracy. However the true data labels, which are found in available dataset, are not available in a production environment, or they can be occasional and possibly hand-crafted. Moreover, the anomalous data samples should be numerous enough to include all the possible anomalies that could happen, or the ones that the IDS should be able to detect.

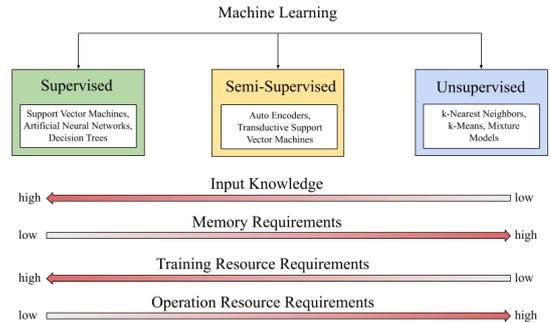

**Fig. 2** – General classification of Machine Learning techniques applied to IDS and their main requirements

The Unsupervised Learning strategy solves the aforementioned problem by not requiring any input label during the training. Unsupervised learning techniques include k-means clustering, Auto Encoders (AEs), Generative Adversarial Networks (GANs), etc. By not requiring the ground-truth labels, Unsupervised Learning techniques are simpler to be deployed in a real world scenario, however they might be more susceptible to noise in the data and the training dataset should be large enough to include many heterogeneous samples. Semi-Supervised learning techniques, which falls in between Supervised and Unsupervised Learning, have also been proposed for the implementation of an AIDS. These techniques use only a small amount of labelled data, achieving a high accuracy and minimizing the task of data labelling of a Supervised Learning system. Overall, Machine Learning-based AIDSs present a wide variety of techniques, enabling the final IDS to be flexible to support the needs of the target deployment environment. They require a small amount of knowledge when compared to expert systems IDSs. Many ML techniques also automatically learn data features, removing the need to hand-craft information for the IDS. Lately, also Recurrent Neural Networks (RNNs), and their specializations like Long Short-Term Memory (LSTMs) and Gated Recurrent Units (GRUs), have been proposed, enabling the IDS to analyze complex and unseen patterns in exchanges between hosts, which is impossible for a knowledge-based or a statistics-based AIDS. On the other hand, ML-based IDSs tend to have a large resource footprint. Depending on the specific technique, the ML model can be computationally expensive, both in memory and CPU and, if using Deep Learning techniques, could also require a GPU. This makes it difficult to run ML-based IDSs on devices with low computa-tional capabilities such as IoT devices. Prediction time required by heavy ML models should be considered. If the targeted system has real-time constraints specialized hardware should be used.

An overview of the most used approaches and their requirements is in Figure 2.

## 3. IDS FOR IOT

As stated in Section 1 IDSs targeting the IoT can be categorized into IoT-specific and IoT-agnostic. An IoT-specific IDS targets devices using a particular communication technology, such as 6LoWPAN, BLE, LoRaWAN etc. This class of IDS should be deployed on the same network of the device. They usually carry out their predictions based on messages sent by the IoT devices leveraging control information of the specific technology, such as checking protocol compliance. On the other hand, IoT-agnostic IDSs do not depend on a particular IoT technology. They utilize information available regardless of which technology is currently used by the devices, such as TCP/IP traffic. This class of IDS is suitable to be used in an edge environment, since it can deal with traffic generated by heterogeneous devices leveraging different communication technologies.

An advantage of an IoT-specific IDS over IoT-agnostic one, is the ability to detect low-level attacks generated on the device-level. On the other hand, a single IoT-agnostic IDS is able to deal with many IoT devices, without the need to deploy an IoT-specific IDS for every communication technology available.

An IoT-specific IDS commonly operates on the network section highlighted in green in Fig. 1, while an IoT-agnostic IDS on the one highlighted in red.

Many IoT-specific IDSs have been proposed for Wi-Fi [40] [41] [42] [43], LoRa [44] [45] [46], ZigBee [47] [48], Bluetooth LoW Energy [49] [50] [51]. These systems are usually expert systems which capture the traffic between hosts and check the compliance of each packet to technology-specific network protocols. Advanced systems, can also detect attacks on the physical network layer (PHY), e.g. jamming. Usually an attacker sending a PHY attack sends bits not following the communication protocol, preventing the data to be readable from an external IDS and making the attack extremely difficult to detect. For example authors in [52] propose an attack on BLE physical layer which selectively jams the signal on specific channels whenever a device tries to connect.

Specific and new security issues arising in the IoT field are the ones related to the network management and operation such as routing, topology control, network maintenance. As for the routing, new protocols for devices with constrained resources have been designed like the Routing Protocol for Low Power Lossy Network (RPL) [53]. The message exchange is based on the Destination Oriented Directed Acyclic Graph (DODAG), which is built by the devices following the protocol, enabling point-to-point and point-to-multipoint communications. Attackers could easily craft malicious packets and disrupt the protocol execution. For this reason many IDSs focus on checking the correct execution of RPL by the connected devices. One of the most widespread attacks is the Rank Attack [54], in which a child node advertises a lower rank value than the real one. The rank value is used to determine which nodes are closer to the root node in a multi-hop scenario, with the rank strictly decreasing from the root to the children. In a Rank Attack scenario, messages could be forwarded along loops, not selecting the most optimized path. Perrey et al. [55] proposed TRAIL, a topology authentication scheme for RPL. The system uses cryptography primitives to enforce the correct rank value of the nodes and thus the correct topology of the DODAG. These cryptography functions are not computationally expensive to run making the system scalable also on low power devices. Chugh et al. [56] studied another kind of attack, the Black Hole attack. Black Hole attack consists in the malicious node dropping all or a fraction of packets routed through it. The authors implemented the attack on various scenarios in 6LoWPAN networks, finding that a reliable detection of the attack is hard. Another family of RPL-based attacks includes Clone ID and Sybil attacks. These attacks consist in malicious nodes copying the identity of legitimate nodes. Zhang et al. [57] studied Sybil categorizing attacks on class based on the ability and the goal of the attacker. They have also presented and categorized defense mechanisms into three main classes named Social Graph-based Sybil Detection (SGSD), Behavior Classification-based Sybil Detection (BCSD) and Mobile Sybil Defend (MSD).

Authors in [58] proposed an system based on Artificial Immune System (AIS). The IDS is distributed among IoT devices, edge nodes and cloud nodes. On the IoT devices, lightweight detectors are deployed. On the edge, alerts are analyzed and processed using Smart Data concepts. Finally, the cloud clusters the data and trains the detectors. In this way the heavy-weight detectors' model training is done on the cloud and only light-weight application of them is carry out by IoT devices.

Verzegnassi et al. [59] proposed a system able to detect Sybil and jamming attacks evaluating the conformity of network parameters across time. The systems passively collects IoT devices network statistics observed by the gateway such as the average signal strength and the average packet rate. These informations are projected onto subspaces of dimensions $D \times L \times N$, where the parameters are respectively the number of devices, the number of considered parameters and the time tick. The output of the algorithm is the value of the conformity of device network parameters through time. An abrupt change in the conformity values can be an indicator of an attack taking place.

As all expert systems, IoT-specific IDSs usually achieve high accuracy and low false positive rates, however they are unable to detect zero-days or unusual usage of the network resources by the hosts. On the other hand, IoT-agnostic IDSs work independently on the communication technology between IoT devices. These IDSs could be deployed on IoT gateways, discarding PHY or MAC layer information, or also in another subnetwork, where they leverage TCP/IP traffic features.

## 4. THE EDGE-ENABLED APPROACH

Edge computing was proposed to enhance the characteristics and the reliability of traditional IoT applications [60], [61] under several aspects. The IoT application can offload computational tasks, storage or management tasks to the edge nodes. Some of the expected quality enhancements include the minimization of the latency, real-time network management and better data management. In this context, also security applications, such as an IDS, could be "migrated" on the edge (see the red box in Figure 1). An IDS could benefit from this transition, having more computational resources available, enabling it to use more complex algorithms, and also more storage capabilities, in order to store systems logs to be later analyzed or to carry out memory-intensive procedures. Also an edge node could offer lower latency than the cloud, which is crucial for real-time IoT applications. Moreover, an IDS deployed on the edge, should be IoT-agnostic, meaning that it does not depend on specific IoT communication technologies. If such an IDS is used, it can deal with many heterogeneous devices using different communication technologies in a unified manner, without having to deploy a single IoT-specific IDS for every subnetwork of devices.

Eskandari et al. [62] developed Passban IDS, an system which is able to apply a protection layer on IoT devices which are directly connected to it. The attacks targeted by the system are TCP/IP-oriented, not including IoT technology-dependent ones, such as Port Scanning, HTTP and SSH brute force and SYN flood. The system does not require intensive calculations and can be deployed also on cheap edge devices and/or IoT gateways, such as Raspberry Pis or equivalent. While the IDS aims to protect devices against a relatively low number of attacks, the system shows a very low false positive rate and high accuracy. A positive note about the system, is that is one of the few fully-implemented IDSs, from the detection algorithm to the alerting system leveraging a web user interface.

Authors in [63] have investigated the identification of malicious edge devices. Indeed, edge devices are privileged for storing and processing data produced by potentially hundreds or thousands IoT devices. For an attacker to gain control over such an edge node, would mean a potential control over the data sent by attached IoT devices. The authors proposed a framework which exploits a two-stage Markov Model, an anomaly-based IDS and a Virtual Honeypot Device (VHD). When an alert is raised by the IDS, it is forwarded to the two-stage Markov Model. The first stage categorizes the specific fog node and the second one predicts whether or not the VHD should be attached to the edge node for which the alert was raised. The VHD stores logs of all attached edge nodes, which can be later investigated by experts. Authors in [64] introduced a system to improve the detection accuracy of an IDS by deploying fuzzy c-means and ANNs in the edge. They compared their approach with classic ANN techniques, and show high accuracy also on attacks with low-frequency.

Hafeez et al. [65] proposed a system to perform anomaly detection at the network edge gateways. The system represents the traffic with features that are agnostic with respect to the IoT communication technology, but only depends on TCP/IP features which can be observed by the edge. The advantage of this approach is that several systems, each one having heterogeneous IoT communication technologies, can be attached to the same IDS. As for the dataset, they have collected IoT data from a real-world test-bed. They have also studied the distribution of the various considered features, and they have observed that the majority of them is well fitted by a heavy-tailed Gaussian. The final anomaly detection is performed through the use of fuzzy clustering. On their custom dataset, they have achieved high accuracy and low false positive rate.

Schneible et al. [66] et al. proposed a framework to perform a distributed anomaly detection on edge nodes. The system consists in deploying Auto Encoder models on several edge nodes positioned in different network regions. The anomaly detection is carried out using the classical Auto Encoder approach, as we described in Section 6.4.1. The system also shows some degree of adaptivity: while deployed the edge nodes update their models based on new observations, identifying new trends in network traffic. An edge node then sends to a central authority its updated model, which aggregates them and sends the updates to the other edge agents. The authors observed that this approach reduces the overhead bandwidth, since the only generated network traffic carries the models of the Auto Encoders instead of all observed data. In this context Auto Encoders were leveraged to detect anomalies as well as an automatic system to extract features compressing observed data, to reduce traffic between edge nodes and the central authority.

While edge nodes have superior computing capabilities with respect to IoT devices, they could not provide resources to perform intensive tasks such as heavy-weight ML model training. Works in literature have foreseen this issue proposing systems which don't require intensive operations. Sudqi et al. [67] have proposed an host IDS running on energy-constrained devices. Sedjelmaci et al. [68] have proposed a more advanced system which makes a trade-off between energy consumption and detection accuracy. Their system is composed by a signature-based IDS, which is more energy efficient but may yield a high number of false positives, and an anomaly-based IDS, which is requires more power to operate but performs a more accurate analysis. During operation, only the signature-based IDS is active. When an alert is raised, it is forwarded to the anomaly-based IDS, which can confirm or discard it. Moreover, the system is formulated as a security game model, where the anomaly-based IDS carry out its predictions based on the Nash Equilibrium. A drawback is that the cloud must be always up and running for the system to work correctly. Anomaly detection techniques could be used not just to detect network intrusions, but could also be used as a means of detecting bugs in devices firmware or deviations from the normal state of a system. In the context of Industrial IoT (IIoT), works have been proposed to detect such anomalies.

Utomo et al. [69] develop a system performing anomaly detection on power grids sensor readings. Anomaly alerts could be used not only as an indication of an illegal intrusion, but also as a means to ensure grid safety preventing failures and blackouts. To perform the anomaly detection, due to the high non-linearity of the readings, an ANN based on Long-Short Term Memory (LSTM) cells is used. LSTM neural networks belong to the family of Recurrent Neural Networks (RNN), a class of ANN architecture which excels in processing data in sequence, such as a sequence sensor readings or a sequence of words in the field of Natural Language Processing (NLP).

Niedermaier et al. [70] found that a single IDS running on the network perimeter could not be able to monitor, capture and analyze all the events. They proposed a distributed IDS based on multiple IIoT agents edge devices and a central unit which unifies the logs produced by them. At its core, the IDS performs anomaly detection using one-class classification techniques: the authors assume that they know the normal behavior of the system, which can be learned by the agents. The IDS is suitable to be run on low-power microcontrollers, since it does not require any intensive calculation. The authors have also developed a proof-of-concept implementation of the system, which is not usually done in similar works.

## 4.1 Device Classification

Recently, efforts have been made to identify and classify IoT devices based on their network traffic fingerprint. Using network packets, classifiers could be built to categorize devices based on their device-class (e.g. motion sensors, security cameras, smart bulb and plugs, etc.) or to learn device signatures such that, if unauthorized devices connect to the network, an IDS is capable of raising alarms, based only on passive readings of the network. Detecting intrusions based only on network traffic is a requirement for IDSs designed to be deployed on the edge or, generally speaking, in another network with respect to targeted devices [71].

Desai et al. [72] developed a feature-ranking system for IoT device classification. The utility of each feature is based on statistical methods. In order to extract features from traffic flows, they considered time windows of 15 minutes, and sub-portion of them, which they named "activity period", corresponding to the time passing from the reception of the first packet to the reception of the last packet device-wise. Based on the class of the device, this activity period can assume different lengths. In their testing, they trained classifiers using both all of the features and only the top-$k$ ranked ones. They found that classifiers trained using only top-$k$ features with $k = 5$, show only a relative $\approx 6\%$ drop in accuracy, meaning a great reduction in computational tasks can be achieved without impacting the accuracy.

Thangavelu et al. [73] proposed a distributed device fingerprinting technique, named DEFT, which recognizes IoT device fingerprint. In the system, the IoT gateways extract features from devices' traffic sessions. These features are then sent to central edge nodes, which gather them and train ML models and classifiers. These classifiers are then sent back to the gateways, which perform the final identification of the device. The system does not need to know in advance traffic signatures of the connected devices, since it can autonomously recognize new devices based on the extracted fingerprint. In particular, when a new device is connected to the network (or an existing device changes its usual traffic e.g. due to a firmware update) the classifier on the gateways marks its traffic as low-probability. In this case the gateway sends the captured features to the edge node. When another device belonging to the same unknown class (i.e. with the same traffic signature) connects to another gateway, this one also sends features to the edge node. Now the edge node is able to clusterize and to identify the new device category. If there is not a second device connecting to another gateway this strategy does not work. The whole system can be controlled as a Software Defined Network (SDN) function. The classification is carried out not packet-wise, which can be expensive in terms of resources, but flow-wise, selecting a fixed time window frame.

Authors in [76] proposed a IoT device classification based on TCP/IP features. The aim of the system is to categorize a device in one of the four different considered classes, which are motion sensors, security camera, smart bulbs and smart plugs. Bidirectional TCP flows are considered for the classification. The selected features are the size of the first $k$ sent and received packets and the $k-1$ inter-arrival times between the first $k$ sent and received packets. Also, the authors have used t-Distributed Stochastic Neighbor Embedding (tSNE) as a means to reduce the dimensionality of the dataset. The presented approach uses only basic features of the TCP/IP stack, nonetheless achieves good accuracy and recall scores, which translates in a classifier not requiring intensive calculations during both training and prediction phases. However also the number of considered device categories is very restricted, indirectly improving

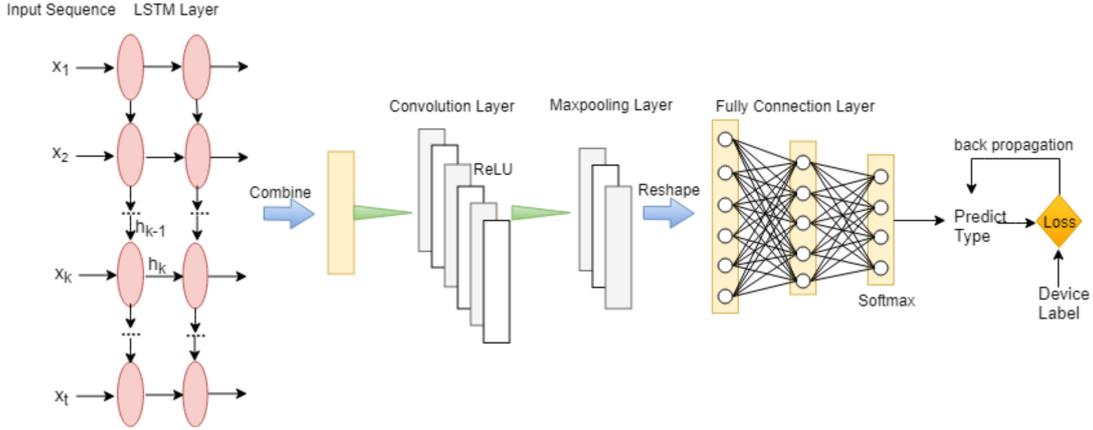

**Fig. 3** – Architecture of the neural network in [74]. The LSTM cells produce an encoding of the network traffic flows. The encoding is given to a convolutional ANN and then to a fully-connected ANN to predict the device label. The network is trained with standard backpropagation/gradient descent algorithms.

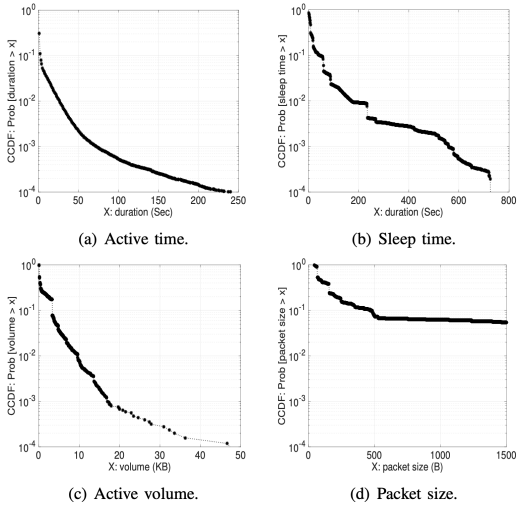

(a) Active time.  (b) Sleep time.
(c) Active volume.  (d) Packet size.

**Fig. 4** – Distribution of some of the features in [75]. The authors collected the raw IoT data over a period of three weeks. The graphs show the Complementary Cumulative Distribution Function (CCDF) of (a) connection duration, (b) sleep time, (c) amount of data shared in the connection, (d) packet size.

the accuracy of the classifier.

Miettinen et al. [77] developed IoT SENTINEL, a system capable of discovering the device-type of each node. The device type includes information such as manufacturer name, model and software version. In order to discover the device-type a machine learning model is built. The idea is to use the device-type against an external vulnerability database in order to identify vulnerabilities and to restrict its communications accordingly. In their testing, they found that their system is effective in identifying device types and has minimal performance overhead. However to identify how to restrict device communication based on the found criticalities is not a trivial task, and was not implemented by the authors. Moreover, if IoT devices with custom firmware are used, the vulnerability database is no longer useful, which makes the system not practical in a real-world general situation.

Bai et al. [74] propose a device classification technique to identify new and unseen devices. This is a novelty when compared to the majority of other works, which in order to recognize a device they must have had some sort of training on that exact device. The classification is done using information streams generated by devices and then using a LSTM-CNN model leveraging time-dependencies of network traffic. First, for each captured packet features such as timestamp, length, and various addresses are saved. Then features are extracted dividing the traffic into time windows of length $T$ seconds. The authors do not specify the value of $T$ that they used in their experiments and it seems to be fixed, non-adaptive. They extracted features differentiating between incoming or outgoing packets and user (TCP, UDP, MQTT, HTTP) or control packets (ICMP, ARP, DNS). Various statistics of the packets are extracted. Finally, the processed data is given to a LSTM network learning an encoding of the data. The LSTM is then attached to a CNN network which learns the final classification. The final architecture is depicted in Figure 3. The network is trained with standard backpropagation algorithms. The achieved results are quite good in accuracy, even if there is room for improvement.

Authors in [75] focused on discriminating if network traffic is generated by IoT or non-IoT devices. The authors didn't use any available dataset, instead they captured network traffic of campus devices over three weeks. They reported some feature distribution among the dataset samples, depicted in Figure 4. To improve the performance of the classifiers, the authors used clusterization techniques for each feature. However it is not clear which algorithm they have used and in which way. The obtained results present good accuracy, however the classifier must be trained with network traffic of each target device, meaning that it is not suited for unseen device classification. This is impractical for systems using a large number of heterogeneous IoT devices.

Bikmukhamedov et al. [78] developed a system which analyzes traffic-flows of IoT network traffic and classifies them using a simple machine learning model. For each flow, features such as statistics on packet lengths and inter-arrival time are considered. Several classification algorithms are then trained, including logistic regression, SVM, decision tree and random forest. To improve the performance also a PCA decomposition of the features is applied. The final classification accuracy is good, considering the relatively high number of considered classes. However, no implementation of flow extraction or feature extraction has been carried out, which could be challenging in a real-world system based on the complexity of the used features.

Hafeez et al. [79] proposed a lightweight technique, named IOT GUARD, to distinguish between malicious and benign IoT traffic, using a semi-supervised approach. Their approach is almost completely unsupervised, but it requires a small portion of labels to be verified by hand, which makes the algorithm technically semi-supervised. It is based on Fuzzy C-Mean clustering (FCM). To improve performances, the authors performed aggregation of same-host and same-service features of devices. This aggregation strategy is not based on packet timestamps, but over the $n$ latest device connections. This brings an advantage since time-based aggregation aggregates features over a definite time, e.g. number of connections made in last $t$ seconds between two devices $A$ and $B$. Time-based aggregation strategy is not suited for detecting attacks where attacker introduces a time-delay between successive connection attempts. In contrast, connection-based aggregation techniques aggregate features over last $n$ connections i.e. out of last $n$ connections made by A how many terminated at device $B$. This technique accommodates the time-delay added to successive connections. The evaluation was carried out using a private dataset which was not made available to the public. The achieved accuracy is good, however practical comparison with other existing solutions or any baseline algorithm weren't made.

# 5. OPEN ISSUES FOR EDGE-ENABLED ARCHITECTURES

The edge network creates new attack surfaces to be exploited by malicious parties. In Figure 1 is illustrated the architecture of an edge-enabled IoT application. Traditional IoT-oriented IDS are placed on the gateway-level or device-level and their focus is to protect against malicious IoT devices. However attacks may target specifically the edge network, making an edge node to become malicious. These could be caused by a remote attack of if the node gets physically tampered. Due to the pervasivity offered by the edge, nodes could be deployed in public areas, which facilitates an attacker to tamper the device. Any attacker in control of an edge node, could alter all the traffic passing through it. They can generate packet streams in the edge pretending to be a legitimate IoT gateway or device, or they can selectively-forward packets of interest and discarding the others. The management of malicious edge nodes was considered in the literature [63], but most of the times it is not an automatic process.

Already existing IDSs could be used and deployed in an edge scenario, however some new challenges arise and hinder the reliability of the Intrusion Detection System. They include:

- *Traffic Encryption.* The IDS can be deployed on IoT gateways or more external edge nodes. If deployed on edge nodes, the traffic it observes is encrypted, assuming that the IoT devices and the cloud communicate through secure protocols. The same could happen if the IDS is deployed on IoT gateways and the IoT devices have a TCP/IP stack, i.e. they can directly communicate with the cloud and the gateway performs routing operations only. Packet encryption means that an IDS is not able to know the contained information and it can only perform operations based on non-encrypted fields, such as TCP/IP headers, timestamp etc.

- *High Resource Variability.* IDSs can leverage several techniques to carry out the detection, which can be highly variable in terms of required computational resources. However also edge nodes show high computational resource variability, which could range from a commodity PC with specialized hardware to a Raspberry Pi. The problem that may arise is that the requested resources for the IDS to work are too high for the edge node which is running the system, which could add communication latency and could block the whole system execution. On the other hand, an edge node that offers much resources costs more, and if the resources are not exploited by the IDS the extra cost is wasted. Edge IDSs should be adaptive to the available resources, using a variety of algorithms requiring different capabilities and selecting them based on the current execution platform.

- *Distributed IDS architecture on Edge/IoT.* Due to the resource variability between IoT and the edge, the execution of IDS for the network edge should be somewhat distributed. A single IDS could be composed of many subsystems which cooperate for the correct working of the system or to improve the detection performance. The cooperation of different subsystems, however, brings distributed systems challenges into the Intrusion Detection System, increasing its complexity.

- *Aggregated traffic.* If the protocol stack of IoT devices and the protocol stack of the edge differ, it could make an observer on the edge, including an IDS, unaware of the source end-device of a packet. This is caused by IoT gateways which receive packets from end-devices using their specific IoT communication technology and craft new packets using the protocols of the network edge, such as TCP/IP. This issue and its aftermath will be illustrated in more detail in section 5.1.

In order to develop communications schemes which are resilient to malicious edge nodes theory of distributed systems could be leveraged, treating edge nodes as potentially byzantine nodes [80] and treating each packet that goes through the edge as a byzantine consensus problem. However, theorems [80] state that, in a non-authenticated and partially synchronous communication scheme, it must hold $N > 3f$, where $N$ is the number of parties and $f$ is the maximum number of tolerated byzantine nodes, in order for a byzantine consensus to be successful. This however would require a transmission of the same packet from multiple edge nodes. Moreover, if the packet was originated from an IoT device, it would require the same device to send the same packet to multiple edge nodes, which is a waste of energy and network resources.

A possible solution to the aforementioned problem is using IPsec [81]. IPsec is a network protocol guaranteeing network layer security, offering authenticity, integrity (AH and ESP modes) and encryption (ESP mode only) to packet header and data. In this way malicious edge nodes are no longer able to craft packets pretending to be a legitimate user of the network. However some issues still persist:

- *IPsec does not protect against traffic rerouting or selective-forwarding.* Attackers could decide which packets to forward and which ones to discard (selective-forwarding). They could also route the packets with additional delay, which can impact the real time characteristic of the IoT application.

- *IPsec fails to guarantee security specification in a physical tampering scenario.* If a device gets tampered, attackers have the possibility to access the private keys of an edge node, compromising the whole IPsec secure communication scheme for the device.

- *IPsec increases fractional overhead.* In a usual IoT application, packets sent from IoT devices are few bytes long, meaning a low ratio of payload data over header data. The use of an additional control header increases even more the payload data, making the communication even more inefficient in terms of fractional overhead.

## 5.1 Aggregated traffic

Another problem is that the edge may not have the possibility to differentiate the traffic flows coming from the IoT devices, in other words it could only observe the aggregated traffic generated by all the devices combined as if it was generated by a single device. This issue verifies whether IoT devices and the edge have different protocol stacks and the gateway has to translate the protocols used by the IoT to the ones used by the edge/cloud. The observation of the aggregated traffic will cause both signature-based and anomaly-based IDS to carry out unreliable predictions.

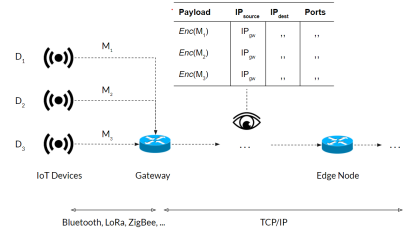

**Fig. 5** – An example of why the edge node may be able to observe only the cumulative traffic, thus being unable to identify the end device who generated the observed packet. Some IoT devices send their data to the cloud. They first communicate to their gateway using their specific IoT communication technology. The gateway then crafts packets which will be sent to the edge and forwarded to the cloud, assuming using TCP/IP as protocols. The packets crafted by the gateway will have the same source IP, possibly the same destination IP (the same application server) and could use the same TCP ports. This causes any observer after the gateway, including an edge node, to be unaware of the devices behind the gateway. The edge node is only able to observe the cumulative traffic, without being able to identify the source device of an observed packet.

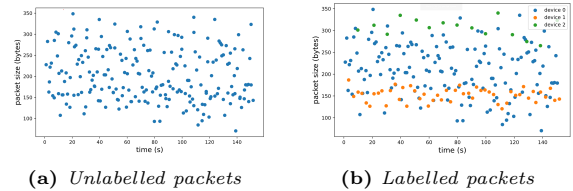

(a) *Unlabelled packets*      (b) *Labelled packets*

**Fig. 6** – Consider three IoT devices. Each device produces packets with its own mean length, its own mean time between them and own variances. In the plots, each dot represent a network packet. As we said in Section 5, the edge is not able to tell which IoT devices are connected and therefore it cannot assign a packet to its most likely IoT producer device. So what the edge observes is an "unlabelled" flow of packets, in fig. 6a, not knowing the source/destination device of a packet. Applying anomaly detection strategies on the cumulative traffic yield poor performance, since too much variance is experienced by the algorithms. In fig. 6b is depicted the same traffic but with packet labelled with its producer device. Applying anomaly detection on the labelled flow, should help algorithms to improve detection accuracy.

Let us consider the scenario depicted in Figure 5. We have IoT devices connected to the gateways with some IoT specific communication technology (BLE, LoRa, etc.) and the gateways connected to the edge and the cloud via TCP/IP. When the IoT devices send data to the cloud, they send a packet to their gateway using their IoT communication technology. The gateway then crafts a new TCP/IP packet and forwards it to the edge and to the cloud. This newly created packet by the gateway will have as source IP address the one of the gateway, regardless of which IoT end-device produced it. Moreover, these packets could have the same IP destination address (same application server) and could use the same TCP ports for every IoT end-device. This causes any observer beyond the gateways, including the edge nodes, to be unable to tell the source device of an observed packet. Being unable to separate the TCP flows, the edge node would regard the observed traffic as it was generated by a single device, since it has no means of knowing which devices are connected beyond the gateways.

The aggregated traffic poses problems for existing IDS, both signature-based and anomaly-based:

- Signature-based IDSs cannot longer isolate packets coming from or going to the same device. This causes the inability to extract patterns from the observed traffic stream, thus making an IDS unable to recognize an attack signature. Methods could be developed to adapt existing signature-based IDSs to solve this issue, for example by mining patterns from the cumulative traffic. However, since the observed traffic is the sum of the various traffic streams generated by each device, there could be cases where a signature could be mistakenly marked as malicious. For instance, let's consider a pattern which is malicious only if generated by a single device (e.g. a particular exchange of messages between it and the server). If two or more devices generate non-malicious messages, it could be that when mining attack patterns, the sum of these flows generates a signature match. This increases the ratio of false positives.

- Anomaly-based IDSs will have to deal with the high variance of the aggregated traffic, since it is presumable for the cumulative traffic to have a higher variance than the traffic flows generated by each single IoT device. To carry out anomaly detection, an IDS has to learn the state of a system in a normal condition i.e. without an anomaly taking place. Then an anomaly is reported when the observed state deviates substantially from the expectation. If the normal state is learned via the cumulative traffic, too much variance could be experienced by the anomaly detection algorithm. The higher variance poses the risk that malicious anomalies are marked as non-malicious oscillations of the expectation, since these oscillations are acceptable given the variance of the normal system state. This increases the ratio of false negatives.

Table 1 summarizes the effects of the cumulative traffic on existing anomaly-based and signature-based IDS.

An example of anomaly detection on the aggregated traffic is illustrated in Figure 6. An anomaly detection algorithm deployed in the edge, should learn the normal system behavior from the cumulative network traffic instead of a device-wise traffic. However the cumulative traffic presents more variance than the traffic split in a device-wise manner, which could drastically impact the performance of the anomaly detection strategy. One first step to improve anomaly detection in the edge, could be to split the cumulative traffic into flows, one for each IoT device. Once this split is done, existing algorithms could be used to learn the normal behavior of the system, not from the cumulative traffic but from the flows of each device. However, this task could not be carried out by an edge node alone, since it doesn't have the knowledge of which IoT devices are connected beyond the gateways.

| IDS Approach | Effects of Cumulative Traffic Observation | Result |
|---|---|---|
| Signature-based | Unable to reliably extract signatures from cumulative traffic | Increase of false positives. |
| Anomaly-based | Anomaly detection algorithm experiences too much variance | Increase of false negatives. |

**Table 1** – Summary of the expected issues that causes the observation of the cumulative traffic on the edge by an IDS. In the case of signature-based IDSs, the system is not able to extract precisely patterns and signatures from the traffic, ultimately increasing the ratio of false positives. In the case of anomaly-based systems, their algorithms would experience too much variance during the learning phase. This will cause an inexact anomaly report with a high ratio of false negatives.

# 6. MACHINE LEARNING TECHNIQUES APPLIED TO IDSS

In this section we discuss some of the widely used Machine Learning techniques applied to IDS and how they could be leveraged by an IDS deployed at the edge. For each technique, we briefly describe the theory behind it and then we illustrate the expected advantages and disadvantages of an IDS based on it. Requirements on computational power, storage capacity and real time response of each technique are highlighted, which make an approach suitable or not for an edge-oriented IDS. A comprehensive overview of the techniques applied to IDS is in Table 2.

## 6.1 Support Vector Machine

Support Vector Machines (SVMs) are a popular ML technique capable of performing classification and regression [82]. In a classification task, training data of size $N$ is given as input in the form $\{(x, y)\}^N, x \in \mathbb{R}^n, y \in \mathbb{R}$. The SVM finds the parameters $w, b \in \mathbb{R}^n$ such that the hyperplane $w \cdot x - b = 0$ partitions the $R^n$ space with instances having same $y$ (labels) falling in the same region. This separation means that the prediction for a new instance $x'$ is given by:

$$y' = \text{sign}(w \cdot x' - b) \quad (1)$$

The hyperplane found by the SVM has the key property of maximizing the separation between the space regions it divides. Let's consider for simplicity $y \in \{-1, +1\}$ and assume a standardized dataset such that hyperplanes $w \cdot x - b = 1$ and $w \cdot x - b = -1$ could be defined. Points $x$ that lay on one of these hyperplanes are named Support Vectors. Geometrically, the distance between these hyperplanes is $\frac{2}{||w||}$ and it is called the margin. The goal of an SVM is to maximize the margin, which is equivalent to minimizing $||w||$. We do not want any point to fall into the margin, so we have to add the constraints $w \cdot x_i - b \geq 1$ if $y_i = 1$ and $w \cdot x_i - b \leq -1$ if $y_i = -1$, which could be written compactly with $y_i(w \cdot x_i - b) \geq 1$. The final SVM optimization problem can be written as:

$$\begin{cases} \min ||w|| \\ y_i(w \cdot x - b) \geq 1, \text{for } i = 1, ..., N \end{cases} \quad (2)$$

and can be solved via Quadratic Programming. However, there could be outliers in the training data having a negative impact on the final hyperplane. For this reason, a soft-margin SVM is usually utilized. This SVM version uses slack variables $\xi_i$ permitting data points to fall inside the margin region at the cost of increasing the error function. The problem can be formulated as:

$$\begin{cases} \min ||w|| + \lambda \sum_i^N \xi_i \\ y_i(w \cdot x - b) \geq 1 - \xi_i, \text{for } i = 1, ..., N \\ \xi_i \geq 0, \text{for } i = 1, ..., N \end{cases} \quad (3)$$

where $\lambda$ is a parameter controlling the weight of the slack variables in the loss function. Until now, we have assumed that the data is linearly separable in its original space, however this could not always be the case. The SVM can solve this problem by applying the so-called kernel trick, transforming inner products $x_i^T x_j$ into $\mathcal{K}(x_i, x_j)$, mapping the data points into a higher-dimensional space. Depending on the task, kernels can be selected to make the data linearly separable in the new space.

To showcase the use of the kernel trick we applied this approach to a real-world IoT dataset derived by a LoRaWAN system provided by an italian service provider (UNIDATA S.p.A.). The deployed LoRaWAN network covers wide geographic areas in Italy and collects a huge amount of IoT data. Specifically, this network currently involves 1862 EDs and 138 gateways. In 2019 it has collected a total of 372,119,877 packets. In Figure 8, each dot is a LoRa packet generated by two real devices. Supposing we want to predict the source device of a packet given its LoRaWAN traffic characteristics, a separation should exist in some feature space. In the plots, each packet is represented in RSSI-SNR space, which are two available data features, with different colors based on the real source device. In Fig. 8a, even if a separation is clearly evident, the data is not linearly separable by a single hyperplane. Training a linear SVM on this data results in the SVM being unable to correctly discriminate between the two classes. To solve this issue we

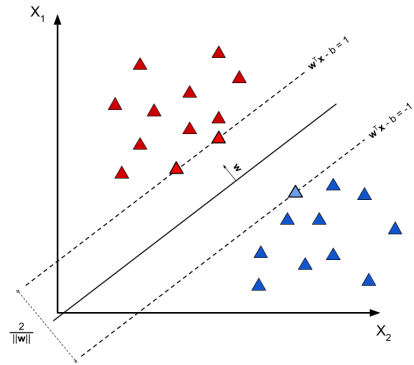

**Fig. 7** – Hyperplane found by the SVM. It maximize the margin $\frac{2}{||w||}$ between data points. The points lying on the hyperplanes $wx - b = 1$ and $wx - b = -1$ are called support vectors.

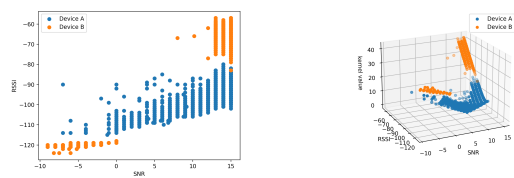

**(a)** *Example of measurements of SNR and RSSI*  **(b)** *Application of a kernel function*

**Fig. 8** – In the figures are plotted the SNR and RSSI of LoRa packets generated by two real devices. In Fig. 8a a linear separation is not possible using a single hyperplane. However, by applying a kernel function we are able to map the data into a higher-dimensional space and to linearly separate the data, as in Fig. 8b. The kernel applied was $K(x, y) = \sqrt{(x-9)^2 + (y+98)^2}$ returning smaller values for data points near $(x, y) = (9, 98)$, which is the centroid of the data points of device A.

can apply a kernel function $\mathcal{K}$ to the data. Let $x, y$ be the two dimensions of the data, then the kernel function creates an additional dimension $z = \mathcal{K}(x, y)$, as illustrated in 8b. In this new space there exists a linear separation of the data, enabling a SVM to learn the separating hyperplane.

In this example we selected an ad-hoc kernel suitable for the available data. A right choice of the kernel is crucial for obtaining good performance from a SVM. However, depending on the data, a good choice for the kernel transformation may be not obvious.

In the context of IDSs, SVMs have been extensively used. With the soft-margin learning strategy and the exploit of the kernel trick, the SVM technique is a powerful and flexible tool for Supervised Learning systems. The SVM requires a fair amount of computational resources during training, not as much of an Artificial Neural Network but more than an instance-based classifier. However, it requires few resources when predicting new values, since the prediction is just a vector multiplication, as can be observed in Eq. *(1)*. This computational efficiency makes a SVM suitable for an online real-time IDS, due to the low added latency to the communication, and enables also devices with low-computational capabilities, such IoT devices, to perform predictions. Simple linear and kernelized SVM approaches have been carried out on public datasets, reaching quite effective detection performance. Authors in [83] have applied basic data preprocessing steps and used a SVM with a Radial Basis Function (RBF) kernel

$$\text{RBF}(x_1, x_2) = \exp\left(-\frac{||x_1 - x_2||^2}{2\sigma}\right) \quad (4)$$

where $\sigma$ is a free parameter. The value of $\text{RBF}(x_1, x_2)$ decreases with the Eucledian distance of $x_1$ and $x_2$, ranging from zero to one (in case of $x_1 = x_2$). For this reason the RBF is often used as a distance measure. The high accuracy of the system, show that high IDS performance could be obtained also with simple approaches. Pervez et al. [84] have proposed a feature-filtering algorithm based on SVM to enhance the performance on NSL-KDD dataset predictions. The algorithm considers a set of input features and trains a first SVM classifier. The algorithm then iterates by modifying the feature space is by removing one of the features via a custom policy. A new SVM is then trained in the new feature space and if the accuracy of the new classifier is greater than the previous one the algorithms continues iterating, otherwise it backtracks. This approach reaches high levels of accuracy and recall (low ratio of false negatives), however it show poor generalization, making the system unsuitable to detect new network intrusions. Chandrasekhar et al. [85] proposed a system composed of the confederated use of Fuzzy C-means (FCM), ANNs and SVMs. The dataset is first partitioned into clusters $K + 1$ via FCM, where $K$ is the number of different attacks that the system will detect and one is the cluster of nominal network traffic. This first partitioning makes the data point variance lower than the whole dataset variance, improving the performance of later applied classifiers. Upon the clustering, one ANN per cluster is applied in order to learn the patterns of the attacks. The final classification is carried out via a binary SVM. The approach is evaluated on publicly available datasets and it shows high values of accuracy and recall for detecting DoS, PROBE, R2L and U2R. However, the system does not perform anomaly detection, which could be useful not only for system security but also to diagnose bugs in the communication protocol. Overall the system requires to be run on devices with high computational capabilities, and may also require specialized hardware such as GPU to train not just one ANN but many ones, one for each attack to be detected by the IDS.

## 6.2 k-Nearest Neighbors

k-Nearest Neighbors (KNN) is an instance-based technique to perform a classification task. It requires labelled samples in the form $\{(x, y)\}^N, x \in \mathbb{R}^n, y \in \mathbb{R}$. The classifier takes as input a data point $x$ and predicts its class to be the most frequent among $k$ training points nearest to $x$, where $k$ is a hyperparameter [86]. Depending on the task, many distance functions can be chosen: some of the most used ones are the Eucledian Distance for continuous data points and Hamming Distance for discrete variables. Correlation coefficients such as Pearson's and Spearman's could also be used as a distance function [87].

KNN-based algorithms have been widely used for IDSs and in particular AIDSs implementations. However, an issue of the "majority voting" strategy is that if the class distribution is highly skewed the prediction performance can be impacted, since there could be a class with a high number of samples which are more likely to "vote" for the classification. As we said, in an IDS context having labelled samples it is not a trivial task, and having balanced labelled samples could be even harder. This could be solved by adding a distance weight in the voting or by using one of the many KNN variants, such as Large Margin Nearest Neighbor or Neighbourhood components analysis [88]. A positive aspect of KNN, and of instance-based techniques in general, for an IDS purposes is the absence of a training phase, which makes KNN suitable for systems whose conditions vary dynamically during time. Suppose we perform anomaly detection on a system and its working conditions change due to a legitimate cause (e.g. seasonality). We may need to rebuild a model for the anomaly detection. If we use methods such as SVMs, ANNs or similar, we have to retrain the whole model, adding down-time to the IDS leaving the system vulnerable. To build a new KNN instance on the other hand, we just need to feed to the algorithm labelled samples and the system would be immediately up and running. One drawback of not building a model, is that predictions may be slower when compared to other ML techniques, impacting the realtime characteristic of the IDS. Exact KNN implementations require $\mathcal{O}(n)$ operations to carry out a single predictions, while more advanced implementations could exploit Locality Sensitive Hashing [89] reducing time to $\mathcal{O}(1)$. LSH is a probabilistic method which hashes inputs $x_1, x_2, ...$ with several heterogeneous hashing functions. For each $x_i$ this technique returns the data points $x_{j \neq i}$ closer to $x_i$. However LSH is a probabilistic method which trades accuracy for computational time (the more hashing function are used the higher the accuracy is) adding error and uncertainty to the final IDS. Another drawback of instance-based algorithms is that the device that carries out the prediction should store the whole dataset (or in case of LSH a hashmap whose size is linear with the dataset) instead of a trained model, such as a hyperplane in case of a SVM. This hinders the possibility of running an IDS on memory-constrained devices. Sharifi et al. [90] proposed a hybrid method involving KNN and k-means Clustering. First, they preprocessed the data via Probabilistic Principal Component Analysis [91] to extract the main 10 components representing the input data. Then they applied k-means clustering to partition the data, assigning to each cluster the most frequent label of the contained data points and, finally, they used these clusters and labels to create a KNN in-stance. They compared the results of their approach with a baseline KNN approach (without the prior clustering) and showed improved performance. Shapoorifard et al. [92] develop a technique improving KNN. Instad of involving only the closest neighbors, they have also considered the farthest neighbors and cluster center. They have combined this approach with a k-farthest neighbors classifier, and show that this hybrid techniques reaches high accuracy and recall. While this approach works well for attack recognition it was not studied how it performs when detecting anomalies in the network. Meng et al. [93] proposed knowledge-based expert system to verify incoming alerts. The system uses multi-tier KNN to filter alarms coming from an already existing IDS. The classification of an alert goes through labelled clusters and expert-made custom rating mechanisms. In their test settings they have merged their system with Snort IDS [94] in such a way to forward all raised alarms to them, evaluating each alert and filtering the false positives. The system achieves a high accuracy score while not loosing too much in recall.

### 6.3 Decision Tree

A Decision Tree is a ML technique which builds a classification tree from input data and uses it to carry out predictions in a short time. Each node of the tree represents a data feature, each branch represent a value of the feature and each terminal leaf represents a possible classification outcome. The Decision Tree model is usually built top-down, selecting feature after feature following a specified policy such as Information Gain (IG). Information Gain gives a measure on how much information is gained on a random variable by measuring another random variable. In the context of Decision Trees, it is defined as the difference in entropy between the prior knowledge $T$ and itself with the value of the attribute $X_i \in T$ given as known:

$$\text{IG}(T, X_i) = \text{H}(T) - \text{H}(T|X_i) \qquad (5)$$

where $\text{H}(T)$ is the Shannon's entropy and $\text{H}(T|X_i)$ is the conditional entropy. Information Gain gives a "ranking" of features, making the tree to first select features with high discriminative potential. The most widely used algorithms to build Decision Trees are ID3, C4.5 and CART [95] and they all leverage Information Gain when creating new tree nodes. The prediction procedure of a decision tree, independently of the algorithm which have generated it, consists in following the tree from the root node, selecting branches by operations of the form if-then-else on the input features. This makes the prediction step of a Decision Tree computationally efficient, without requiring specialized hardware as other ML techniques. The main issue of using a Decision Tree is overfitting training data. While overfitting is a common issue in all ML-based techniques, Decision Tree particularly suffer from it, since it only compares features independently without performing any association. To avoid overfitting, techniques such as post-pruning could be used. A pruning procedure takes a tree and removes some of its internal nodes, trying to improve model generalization by removing some training set-specific checks. It is not a trivial task to decide a threshold on how much pruning should be applied to a decision tree. Alternating pruning and evaluation on test set, may not be a good strategy to follow, since using the test set to build the model makes the model itself biased towards the test set. In order to obtain non-biased test results, the dataset should be partitioned in more than two sets, e.g. using $k$-fold cross validation.
Decision Trees-based techniques may be the most widely used for existing IDSs [96]. Their implementation simplicity and their fast building and prediction speeds make them suitable for most of the IDSs needs. However, the need for more training data to avoid overfitting, is an issue for IDSs, since the data could be labelled by hand. Another issue is that it is difficult to model complex relationships between features. In the prediction phase, each feature is treated independently from the others, which could cause an incorrect classification of some particular inputs. A possible solution to this problem could be a first data processing data via PCA. Another important issue regarding the application of Decision Trees to an IDS, is the impossibility to identify patterns which extend among different data points. Indeed, Decision Trees treat each input independently from the others and do not maintain any inner state of the previous predictions. This means that an IDS leveraging only Decision Trees cannot identify patterns in packets exchanged by the hosts, which prevents the IDS to detect a wide variety of potentially malicious messages. For this reason Decision Trees are not well suited for identify zero-day attacks and performing anomaly detection, although some literature leveraged Decision Trees enhanced with other ML technique [97], [98], [99]. Malik et al. [100] proposed an Intrusion Detection System based on Decision Trees with a technique based on Particle Swarm Optimization (PSO) to prune the tree. Both single-objective and multi-objective pruning were tested. They found that the multi-objective strategy is more suitable to reduce the whole size of the tree, while the single-objective strategy enables the model to achieve a much higher generalization. Rai et al. [101] developed a technique to build a Decision Tree-based IDS handling issues of split value and feature selection. The building algorithm is based on Ross Quinlan's C4.5 and the selection of the feature to split on Information Gain. However the split is decided not ranking the attributes for each node, but by taking the mean. This makes the split to be unbiased with respect to the most frequent attribute values. Azad et al. [102] propose a hybrid IDS based on Decision Tree and Genetic Algorithm (GA). It presents a solution to the problem of the Small Disjoint in a decision tree. The Small Disjoint problem arises when nodes closest to the leafs discriminate only a small number of instances, leading to overfit the training data. The genetic algorithm is used to improve the coverage of those rules which are cope with the problem of the Small Disjoint. The system consists of a first module generating the rules and a second one optimizing them.

## 6.4 Artificial Neural Network

Artificial Neural Networks (ANNs) are mathematical models which can approximate any continuous function $f : \mathbb{R}^n \to \mathbb{R}^m$. Neural Networks are usually composed by many neurons arranged in layers. A neuron takes the outputs of all the neurons in previous layer and perform the following operation

$$o(x) = \sigma \left( \sum_i w_i x_i + b \right) \quad (6)$$

where $x_i$ are the outputs of the previous neurons, $w_i$ are the weights of previous neurons, $b$ is a real valued parameter called bias and $\sigma$ is a non-linear activation function. In the case of fully-connected networks, there exists a weight between each neuron belonging to a layer of size $h_1$ and each neuron belonging to the next layer of size $h_2$, which brings the total number of weights equal to $h_1 \cdot h_2$. Convolutional Neural Networks (CNN) on the other hand, exploit parameter sharing to reduce the number of parameters [103]. Their parameters are contained in kernels performing convolutions over the input. Many kernels are usually applied on a single layer, however the number of parameters is orders of magnitude lower than if using a fully-connected network. This brings a performance improvement in both training and prediction steps. Moreover parameter sharing makes the network learn the target function such that it is invariant to forms of deformation such as translation, scaling or tilting, which is especially useful when dealing with image or text recognition.

Mathematical theorems belonging to the family of Universal Approximation Theorems [104] guarantee the possibility that a continuous function could be approximated by a neural network. One of the first ones was Cybenko's theorem [105], stating that any function $f$ can be approximated by a fully-connected neural network with one hidden layer.

*Cybenko's Universal Approximation Theorem*
*Let $f : [0,1]^n \to [0,1]^m$ be a continuos function. Let $x \in [0,1]^n$, $w_j^1 \in \mathbb{R}^n$, $w_j^2 \in \mathbb{R}^m$, $b_j \in \mathbb{R}$ and $\sigma$ be a sigmoid function $\frac{e^x}{e^x+1}$ Consider finite sums of the form*

$$g(x) = \sum_{j=1}^{h} w_j^2 \sigma \left( \left( w_j^1 \right)^T x + b_j \right)$$

*There exist $w_j^1, w_j^2, b_j$ for $j = 1, \ldots, n$ such that*

$$|f(x) - g(x)| < \epsilon$$

*for any $\epsilon > 0$ and for some value of $h$.*

Cybenko's theorem states that any continuous function could, in theory, be approximated by a neural network with only one hidden layer. However, it does not give any bounds on the size $h$ of the hidden layer and, more importantly, it does not tell how to train the network i.e how to find those values of $w, b$. One of the most widely used training algorithms is the Back-propagation [106] [107], consisting in calculating the derivative of the error with respect to the network weights and them descending the error function gradient. Let $E$ be the error function, $o$ be the output of a neuron and $net$ be the non-activated output. Then, using the chain rule, we can write the derivative of the error w.r.t. a weight $w_{ij}$ of a neuron

$$\frac{\partial E}{\partial w_{ij}} = \frac{\partial E}{\partial o_j} \frac{\partial o_j}{net_j} \frac{net_j}{\partial w_{ij}} \quad (7)$$

Depending on the activation functions, closed forms of each of the three terms in Eq. *(7)* could be calculated. Finally, the weight $w_{ij}$ is updated by gradient descent [108]:

$$w_{ij} \leftarrow w_{ij} - \eta \frac{\partial E}{\partial w_{ij}} \quad (8)$$

where $\eta$ is the Learning Rate hyperparameter.
Neural Networks, especially CNNs, have been used for IDS to identify attacks from network traffic. Wang et al. [109] proposed a CNN-based approach to detect malicious network traffic. To better exploit the generalization offered by a CNN, they have first developed an encoding scheme which transforms a data sample into a one-channel 2D image. This encoding does not depend on any hand-crafted knowledge nor application-specific information, thus is applicable on the raw packets in any IDS. Their system achieves fair accuracy, even if there is room for improvement. Wang et al. [110] proposed a 1D CNN for classification of encrypted traffic. The ability of performing machine learning operation of the encrypted traffic, enables the IDS to potentially be deployed in on another network of the target system, where the incoming traffic to analyze is almost always encrypted due to the use of TLS or other encryption schemes. We will see that this is one of the issues of deploying an IDS in the Edge in Section 3 and 5. Regarding the performance of a CNN-based approach, despite a CNN's higher number of trainable parameters when compared with other machine learning techniques, the accuracy of vanilla CNNs are not higher [29]. During the training phase, batches of training data samples are feed-forwarded into the network and gradients, for each sample and for each weight, are calculated and kept in memory. This means that training networks requires an immense amount of computation, and for this reason they are almost always trained by means of specialized hardware, such as high-end GPUs, capable of performing massively parallel floating-point calculations. On the contrary, after the training the network can perform the predictions by using only the CPU. In the context of IDSs, this means that a network could be trained offline (also off-site e.g. using remotely a cluster of GPUs of a cloud provider) without the need of installing specialized hardware inside the system targeted by the IDS. While CNNs have been extensively applied for attack recognition, they have been hardly used as a means of detecting network anomalies [111]. In theory, it could be possible to train a CNN on data having labels "normal" and "anomaly", however the wide variety of anomalies attacks requires a huge training set that includes all possible anomalies that can be detected by the system. We will see in Section 6.4.1 that the standard way of detecting anomalies with a neural network is via the use of an Auto Encoder, a particular network architecture that makes easier the task of one-class classification.

### 6.4.1 Auto Encoder

Auto Encoders (AEs) are a special ANN architecture which is able to learn an encoding of the input data using unsupervised learning. This encoding can then be used for dimensionality reduction, noise cancelling and anomaly detection. They consist in two neural networks, namely the encoder and the decoder. The encoder is a network with $n$ inputs and $m \ll n$ outputs, while the decoder has a "mirrored" architecture with $m$ inputs and $n$ outputs. The training of an AE is unsupervised, with the encoder learning $E: \mathbb{R}^n \to \mathbb{R}^m$ and the decoder learning $D: \mathbb{R}^m \to \mathbb{R}^n$ such that $||x - D(E(x))||^2$ is minimized. The vector $E(x) \in \mathbb{R}^m$ represents the encoding of $x$ in a lower space. Many variations of an Auto Encoder exists offering regularization properties on the encoded space, which a vanilla a AE does not [112]. The most widely used include:

- *Variational Auto Encoder* [113], which adds some sort of continuity to the encoding. In other words if $D(E(x)) \approx x$ then $D(E(x) + \epsilon_1) \approx x + \epsilon_2$. It does this by adding to the classic reconstruction loss $||x - D(E(x))||^2$ a term which pushes the encoding space to follow a normal distribution.

- *Denoising Auto Encoder* [114], which learns to reconstruct a sample from a corrupted input. During training, samples $x$ are artificially corrupted via a stochastic process $\hat{x} \sim q(\hat{x}|x)$. The reconstruction should be $D(E(\hat{x})) \approx x$, forcing the encoder/decoder to learn to denoise $\hat{x} \to x$.

- *Sparse Auto Encoder* [115], which presents an architecture with a higher number of neurons in the bottleneck. However, by adding a penalization term to the loss function, only few of them are active during the encoding, depending on the input. The obtained sparse encoding is reported to increase the performance in classification tasks [116] [115] [117].

Anomaly detection belongs to a family of machine learning tasks named one-class classification, in which a ML model is trained to perform a binary prediction on whether or not an input sample belongs to a particular class, and Auto Encoders are widely used for this goal [118] [119] [120]. The usual anomaly detection strategy to follow when using an AE, is to train the model only on non-anomalous samples. Then, when the system is deployed, it encodes and decodes real samples. When the reconstruction error gets higher than some threshold $||x - D(E(x))||^2 > \delta$, an anomaly alert is raised. Due to the unpredictable variability of the system, vanilla AE should be avoided in favor of variants such as Variational AE. How to precisely set the threshold $\delta$ however is not trivial, and often depends on how much the state of a system can change over time. Moreover, if system conditions vary too much, it could be required to train from scratch the AE, which brings all the downsides of heavy-weight ANN-based systems. On the other hand, an IDS trained in a unsupervised way is a huge advantage, since all the labelling procedure of the dataset could be skipped. Normal data samples could be gathered automatically during periods of time which are intrusion-free by assumption of using another IDS. Abolhasanzadeh et al. in [121] proposed a system to improve the detection performance of an IDS. They used the encoding of a vanilla AE as a means to reduce dimensionality. They performed the final classification via the use of an ANN. They reported a substantial decrease in the prediction latency, which is an issue for ANN-based IDSs. In the same work, they have also compared their AE-based dimensionality reduction with PCA, factor analysis and non-linear kernelized PCA [122], with their method achieving higher accuracy than any other technique. Aminanto et al. [123] proposed an AE-based IDS specifically to detect Wi-Fi impersonation attacks. They performed feature extraction using stacked Auto Encoders, a particular class of AE which has multiple encoding layers [124] [125]. After feature extraction phase, they have used ML techniques such as SVM, ANN and Decision Tree to give a feature weighting. The final classification, considered the weighted features is performed via an ANN.

### 7. CONCLUSIONS

This work presents Intrusion Detection Systems for IoT, both under the architectural perspective and under the methodologies that are used to let them capture anomalies and cyber attacks. As for the architectural perspective, while traditional IoT IDS are deployed at device-level or at gateway-level the strong interest in having edge computing solutions offers new attack sides to be exploited by malicious parties. We discussed these new issues and present solutions that have been introduced to face them. New IDSs, specifically designed for the edge, are addressed. We then focus on the adoption of Machine Learning techniques that are nowadays leveraged by IDSs. For each technique, we described the theory behind it and then we illustrated the expected advantages and disadvantages. Requirements on computa-tional power, storage capacity and real time response of each technique are highlighted, which make an approach suitable or not for an edge-oriented IDS.

## AUTHORS

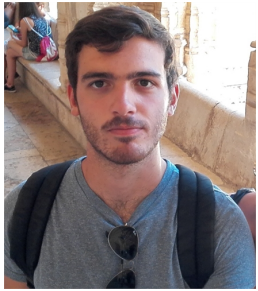

**Pietro Spadaccino** received his B.S. in Computer Engineering and his M.S. in Computer Engineering from University of Rome Sapienza, Italy in 2018 and 2020, respectively. He is currently pursuing a Ph.D. degree in Information and Communication Technologies Engineering at Sapienza. His research interests focus on Internet of Things, LoW Power Area Networks and network security.

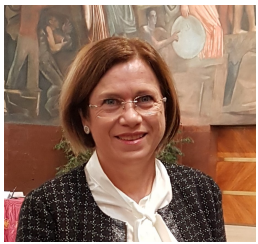

**Francesca Cuomo** received the Ph.D. in Information and Communications Engineering in 1998 from Sapienza University of Rome. From 2005 to October 2020 she was Associate Professor and from November 2020 she joined "Sapienza" as Full Professor teaching courses in Telecommunication Networks. Prof. Cuomo has advised numerous master students in computer in computer engineering, and has been the advisor of 13 PhD students in Networking. Her current research interests focus on: Vehicular networks and Sensor networks, Low Power Wide Area Networks and IoT, 5G Networks, Multimedia Networking, Energy saving in the Internet and in the wireless system. Francesca Cuomo has authored over 158 peer-reviewed papers published in prominent international journals and conferences. Her Google Scholar h-index is 31, >3947 citations. Relevant scientific international recognitions: 2 Best Paper Awards. She has been in the editorial board of Computer Networks (Elsevier) and now is member of the editorial board of the Ad-Hoc Networks (Elsevier), IEEE Transactions on Mobile Computing, Sensors (MDPI), Frontiers in Communications and Networks Journal. She has been the TPC co-chair of several editions of the ACM PE-WASUN workshop, TPC Co-Chair of ICCCN 2016, TPC Symposium Chair of IEEE WiMob 2017, General Co-Chair of the First Workshop on Sustainable Networking through Machine Learning and Internet of Things (SMILING), in conjunction with IEEE INFOCOM 2019; Workshop Co-Chair of AmI 2019: European Conference on Ambient Intelligence 2019. She is IEEE senior member.

Table 2 – An overview of expected advantages and disadvantages of an IDS leveraging a specific Machine Learning technique.

| Machine Learning Technique | References | Expected Intrusion Detection System Advantages and Disadvantages |
|---|---|---|
| SVM | [83], [84], [85], [126], [127] | **Advantages**: Requires few resources when predicting new values and low memory to store the trained model. Suitable to be run by IoT devices. Can be applied to real-time applications.<br>**Disadvantages**: Demands a high amount of resources during training, fine-tune kernel transformation to adapt to the data. |
| k-NN | [90], [92], [93], [128], [129], [130] | **Advantages**: Non-parametric approach, deletes IDS downtime related to training, adaptive to system state changes and suitable to carry out anomaly detection. For small datasets it requires few resources to be run. For larger datasets Locality Sensitive Hashing could be used carrying out predictions with $\mathcal{O}(1)$ complexity.<br>**Disadvantages**: Non-parametric approach, requires the entire dataset to be stored in memory, not just a trained model. Prediction complexity scales with the size of the dataset. Locality Sensitive Hashing adds new complexity to the implementation and a degree of error and uncertainty to the final IDS. |
| Decision Tree | [100], [101], [102], [131], [132] | **Advantages**: Few resources in both training and prediction. Widely used in existing IDS. Low memory requirements to store the trained model. Suitable for real-time applications. PCA could be used to model complex feature relationships.<br>**Disadvantages**: Difficult to model complex feature relationships, since each feature is treated independently from the others. PCA can partially solve this issue, however it adds new complexity and demands resources to be computed, which may or may not hinder the possibility to run the IDS on low powered devices, depending on the size of the dataset. |
| ANN | [109], [110], [133], [134] | **Advantages**: Able to learn complex decision functions for several traffic scenarios. Trained ANN models could be run on low-powered devices, if enough memory is available. Suitable for the precise classification of a network attack. Specialized ANN, such as Recurrent Neural Networks, could be used to carry out prediction based on complex traffic patterns.<br>**Disadvantages**: Training requires a high amount of computational resources and, often, specialized hardware such as GPUs. A poor design of the network architecture could hinder the performance of the final IDS. If the goal is to perform anomaly detection, a dtaset of normal and anomalous traffic should be provided, which is not trivial to be built. |
| Auto Encoder | [121], [123], [135], [136] | **Advantages:** Specialized neural network with a "bottleneck" architecture. Suitable for finding an encoding of the input data. Recurrent auto encoders could be used to encode complex traffic patterns. If the goal is anomaly detection, a dataset composed of oly normal traffic could be provided, then carrying out predictions based on the reconstruction loss of new unseen inputs (one-class classification).<br>**Disadvantages:** Requires careful selection of the architecture. Resource-demanding operation. If performing anomaly detection, the system could be retrained every time the normal system conditions change. |